\documentclass[12pt,a4paper,final]{iopart}
\pdfoutput=1
\usepackage{iopams}
\usepackage{graphicx}

\usepackage[breaklinks=true,colorlinks=true,linkcolor=blue,urlcolor=blue,citecolor=blue]{hyperref}
\usepackage[collision]{chemsym}
\usepackage{xspace}

\begin{document}

\setcounter{footnote}{0}
\newcommand{\AlOx}{\Al\O$_\mathrm{x}$\xspace}
\def\Lkin{$L_\mathrm{kin}$}
\def\Deg{$^\circ$C}

\title{Aluminium-oxide wires for superconducting high kinetic inductance circuits}

\author{H Rotzinger$^{1}$, S T Skacel$^1$, M Pfirrmann$^1$, J N  Voss$^1$, J  M\"unzberg$^1$, S  Probst$^1$, P  Bushev$^1$, M P  Weides$^1$, A V  Ustinov$^{1,2}$ and J E  Mooij$^{1,3}$}
\address{$^1$Physikalisches Institut, Karlsruher Institut f\"ur Technologie, Wolfgang-Gaede-Str. 1, 76131 Karlsruhe, Germany}
\address{$^2$Russian Quantum Center, 100 Novaya Street, Skolkovo, Moscow region 143025, Russia}
\address{$^3$Kavli Institute of NanoScience, Delft University of Technology, Lorentz weg 1, 2628 CJ Delft, The Netherlands}

\ead{hannes.rotzinger@kit.edu}

\begin{abstract}
We investigate thin films of conducting aluminium-oxide, also known as granular aluminium, as a material for superconducting high quality, high kinetic inductance  circuits. The films are deposited by an optimised reactive DC magnetron sputter process and  characterised using microwave measurement techniques at  milli-Kelvin temperatures. We show that, by precise control of the reactive sputter conditions, a high room temperature sheet resistance and therefore high kinetic inductance at low temperatures can be obtained. For a coplanar waveguide resonator with 1.5\,k$\Omega$  sheet resistance and a kinetic inductance fraction close to unity, we measure a quality factor in the order of 700\,000 at 20\,mK.  Furthermore, we observe a sheet resistance reduction by gentle heat treatment in air. This behaviour is exploited to study the kinetic inductance change using the microwave response of a coplanar wave guide resonator. We find the correlation between the kinetic inductance  and the sheet resistance to be in good agreement with theoretical expectations.

\end{abstract}


\vspace{2pc}

\maketitle

\section{Introduction}

Superconducting wires with a high kinetic inductance have been an active research topic in the past two decades in several different research areas \cite{Ra93}. From a technological perspective, the wires offer the possibility to overtop the geometric inductance by orders of magnitude and therefore allow, for instance, for very compact microwave resonator structures \cite{Av13}. Also, the steep temperature dependence in the vicinity of the superconducting transition temperature can be used as a sensitive thermometer in superconducting low-frequency \cite{Au93} and microwave particle detectors \cite{Da03}. An emerging new field is the use of superconducting high kinetic inductance wires for parametric amplification of microwave signals \cite{Vi15}, e.g. for superconducting qubit measurements. In this field, a wire with a large kinetic inductance may replace also arrays of Josephson junctions, which are often used to enhance the inductance of a circuit.  
More fundamentally, a high kinetic inductance is among other aspects a mandatory ingredient for quantum phase slip measurements \cite{La01,Mo06,Ar08,As12,Mo15} and also in the context of experiments with quantum phase transitions \cite{RS13,Ov13}. 

In this paper, we explore a new possibility of using aluminium oxide (\AlOx) for obtaining a high kinetic inductance in the  superconducting state.  The material has been researched as a \emph{granular} superconductor, often referred to as \emph{granular aluminium}, since the late 1960's \cite{Ab66,Co68,De73, Ab76, Zi78} until now \cite{Ba13,Ba14}, with a large body of publications on the film properties, transport measurements and even absorptive microwave measurements \cite{Da69,St85,Su89}. However, the available high kinetic inductance of structured superconducting wires was not of interest.

Conventionally, the previously studied films have been prepared in a low pressure oxygen atmosphere and by thermal evaporation of pure aluminium. By employing this method, the room temperature electrical sheet resistance can be varied on a large scale from low ohmic (m$\Omega$), over very high ohmic (k$\Omega$) and to electrically insulating by simply adjusting the oxygen partial pressure.  The obtained films cannot be considered microscopically "homogenous" due to the intrinsic granularity. The grain size in the high ohmic regime was determined with transmission electron microscope imaging to about 4\,nm.\setcounter{footnote}{0}\footnote{The findings in Ref.~\cite{De73} are in good agreement with the grain size measured on our sputter deposited films.} 

The large range of possible sheet resistances stem from the variable thickness of a thin aluminium-oxide layer that is covering the individual aluminium grains. From this perspective the films can be seen as a network of Josephson junctions mediating the superconducting transport, with a very sensitive dependence on the insulating layer thickness and thus the implanted oxygen content \cite{De73}.

In the following, we are purely interested in superconducting films at temperatures much below the transition temperature of about 1.6\,K-1.8\,K (20\,nm thick films). To obtain a superconducting film, however, the normal state sheet resistance $R_\mathrm{n}$ should be roughly below the quantum resistance $R_\mathrm{q} = h/4e^2 = 6.45\,\mathrm{k}\Omega$ \cite{Ha89,Ja89}. 

For functional wires with a large kinetic inductance also the film quality and $R_\mathrm{n}$ reproducibility in the sub $R_\mathrm{q}$ range are  key aspects. Instead of the thermal evaporation technique, we have focused on controlling  $R_\mathrm{n}$  by employing a reactive DC magnetron \emph{sputtering process} in an oxygen atmosphere. This method is described in section \ref{SampleFab}.  In section \ref{ThermalTreatment} we discuss the $R_\mathrm{n}$ dependence on thermal annealing at room temperature. The results are used to study the kinetic inductance of \AlOx wires structured as microwave coplanar waveguide resonators at mK temperatures (section \ref{cpw_measurement}). 

\section{Theoretical background}

Similarly to an argumentation found, e.g. in Refs.~\cite{An10, Ti04}, the kinetic inductance  $L_\mathrm{kin}$ of a superconducting wire can be derived using the BCS theory. In the low frequency limit ($h f \ll k_\mathrm{B}T$), the Mattis-Bardeen formula for the complex conductivity in the local, dirty limit can be written in terms of the ratio of the imaginary conductivity $\sigma_2$ to the normal state conductivity $\sigma_\mathrm{n}$  

\[ \frac{ \sigma_2}{\sigma_\mathrm{n}} = \frac{ \pi \Delta(T)}{h f} \tanh{ \frac{\Delta(T)}{2 k_\mathrm{B} T}} \]
where $\Delta(T)$ is the superconducting energy gap. Using the BCS relation $\Delta(0) = 1.76 k_\mathrm{B} T_c$, the expression simplifies at temperatures much below $T_\mathrm{c}$ to 
$ \sigma_2/\sigma_\mathrm{n} =  \pi \Delta(0) / h f  =  1.76  \pi k_\mathrm{B} T_c/ h f$. The imaginary component of the impedance is due to kinetic inductance $L_\mathrm{kin}' = 1/ 2 \pi f \sigma_2$.  At $T \ll T_c$ the total kinetic inductance $L_\mathrm{kin,tot}$ of a wire with the length $l$ and the width $w$ can then be written as the product of the sheet inductance $L_\mathrm{kin}$ and the number of squares $N=l/w$ 
\begin{equation}\label{Lkin_R}
L_\mathrm{kin,tot} = N L_\mathrm{kin} = 0.18 N \frac{\hbar R_\mathrm{n}}{k_\mathrm{B} T_c},
\end{equation}
where $R_\mathrm{n} = 1/\sigma_{2}$ is the normal state sheet resistance. This handy formula is used at various places in this paper.

We use a coplanar waveguide (CPW) resonator made from \AlOx at $T\ll T_c$ to obtain $L_\mathrm{kin}$ by measuring the resonance frequency. These data will be compared with $L_\mathrm{kin}$ evaluated from the sheet resistance. Neglecting the  frequency shift induced by the coupling capacitors, the resonance frequency of a $\lambda/2$ CPW resonator is given by 
\begin{equation}\label{freq_LC}
f_n= n/2l\sqrt{L C},   
\end{equation}
where $n$ is  an integer harmonic number. The center conductor inductance $L$ and the capacitance of the center conductor to ground $C$ are per unit length.  $L$ constitutes of $L_m + g_\mathrm{kin} L_\mathrm{kin}$,  the sum of the "geometric" or "magnetic" inductance $L_m$ and the kinetic inductance $L_\mathrm{kin}$. The factor $g_\mathrm{kin}=\alpha  N/l =\alpha /w$ is discussed in section \ref{lambda_2}, where $\alpha$ is in the order of unity. Using the conformal mapping technique, $L_m$ and $C$ are analytically found to be  $L_m = \mu_0/4 \ K(k') / K(k)$ and $C = 4 \epsilon_0 \epsilon_\mathrm{eff} K(k)/K(k')$, where $K$ is the complete elliptic integral of the first kind with the moments $k = w/(w+2s)$ and $k' = \sqrt{(1-k^2)}$ and $s$ being the distance between the center conductor and ground \cite{Yo95,Si01}, see also Fig.\,\ref{setup_l4res}.

\section{Sample fabrication and resistance measurements}
\label{SampleFab}
The CPW resonators are fabricated in three steps, first, a 20\,nm \AlOx thin film is grown by DC magnetron sputter deposition on to an intrinsic or SiO$_2$ passivated 20\,$\times$\,20\,mm$^2$ \Si\  substrate, details of this process are given below. The second step employs optical lithography to define the resonator structures on top of the \AlOx thin film. Here we use the common 1\,$\mu$m thick Clariant AZ-5214E optical resist and an UV-light mask aligner. The optical resist pattern is subsequently used as etch mask in a Cl/Ar 10:1 ICP/RIE plasma at 200\,W/100\,W etch process for 95\,s. After that, the resist mask is removed and the substrate diced into 5\,$\times$\,5\,mm$^2$ chips.  
    
\subsection{Reactive DC sputter process}\label{sputter}

A common way of growing ex-situ non-insulating \AlOx thin films \cite{Ab66,Co68,Ba13,De73} employs the thermal evaporation of aluminum in an oxygen atmosphere at about $1.0\times10^{-5}$\,mbar pressure. Initial experiments with a thermal e-beam evaporator and this technique showed an insufficient reproducibility of the sheet resistance at a given partial pressure of oxygen, especially for consecutive evaporations in the same chamber and with the same aluminium source. As a main reason for this, we identified the oxygen contamination of the hot, electron beam heated aluminium source. This limitation we overcome by using an \Ar\  plasma DC and also a pulsed DC magnetron sputtering process. Here, the source is usually water cooled and it getters the injected oxygen ions only at the "cleaned" surface. All films are deposited at room temperature.

We employ a home-made two-chamber sputter tool, having a base pressure of  $<6\times10^{-8}$\,mbar, with three 2" sputter targets, one equipped with 6\,N purity Al. Prior to the sputter deposition, the samples are cleaned in the load lock using an RF \Ar\ plasma for 2\,min at a power of 20\,W, mainly to remove water from the surface. Vital for reproducible results  is the cleaning of the sputter target in the main chamber, which we do for at least 5\,min in a pure Ar plasma at 100\,W power. The reactive sputter process is then started by injecting  2.5\,sccm \Ar/\O$_2$ 9:1 mixture using a 10\,sccm mass flow controller into the main chamber. The reactive plasma is stabilised at 100\,W power and Ar flow of $\sim$40\,sccm for approximately 1\,min before a shutter is opened and the film is grown on a $\sim$60\,rpm rotated sample for 3.5\,min (rate $\approx$ 5.7\,nm/min). We found the stability and resolution of the mass flow controllers as well as the pressure control to be important for reproducible results. The above parameters for the sputter process gave good results. Furthermore, we found several combinations of DC sputter powers and \Ar/\O$_2$ gas flow / pressure yielding similar results. 

\begin{figure}[h!]
 \centering
 \includegraphics[scale=0.4]{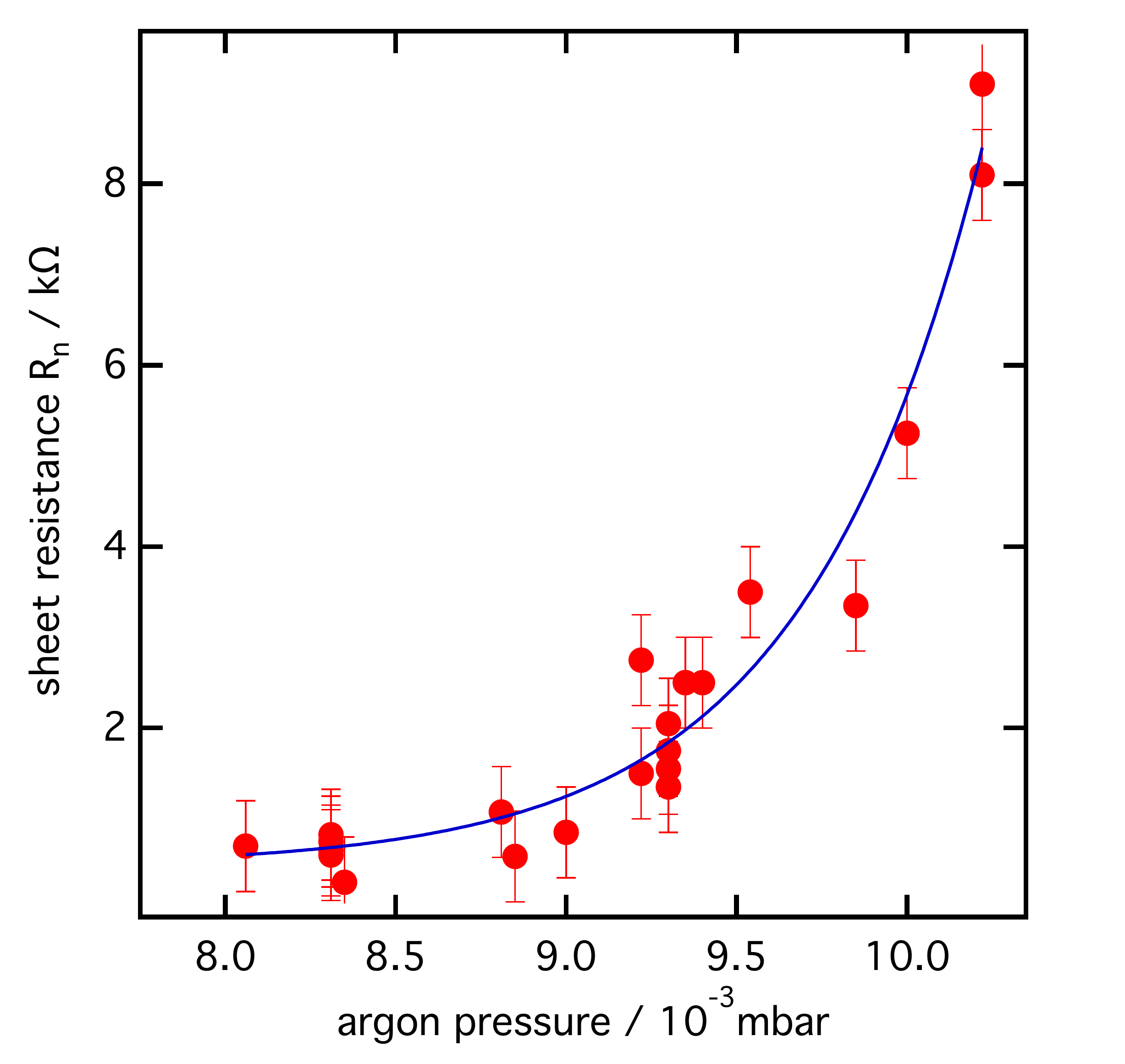}
 \caption{\label{R_sq_vs_P} \AlOx sheet resistance depending on the \Ar\ partial pressure during the sputtering process. The oxygen flow is kept constant for all films. The solid line is a fit to the data, details are given in the text.}
\end{figure}

The target sheet resistance of  \AlOx films depends on the amount of implanted oxygen in the film \cite{Ab66,Co68,De73,Zi78} and, therefore, the ratio of the growth rate to the oxygen partial pressure. 
The sputter tool lacks a direct measurement of the growth rate e.g. through a quartz crystal monitor. Therefore, we use the \Ar\ partial pressure as a control parameter and keep the oxygen flow constant. We are interested in the sheet resistance value range from 0.1 to 10\,k$\Omega$. In this range, the process parameters for $R_\mathrm{n}$ are well described by the following empirical formula $R_\mathrm{sq} = R_0 + R\,\exp[(P-P_\mathrm{off})/P_0]$ with  $P$ the \Ar\ partial pressure. Least square fitting of data of various samples, see Fig.\,\ref{R_sq_vs_P}, yields to $R_0 = 0.5\,\mathrm{k}\Omega$, $R = 0.11\,\mathrm{k}\Omega$, $P_0 = 0.52\times10^{-3}$\,mbar, keeping $P_\mathrm{off}$ fixed to $8\times10^{-3}$\,mbar. 
Not included here is a linear shift in the sputter conditions due to the target erosion, corrected over larger time scales. This shift was evaluated experimentally by using a test sample, prior to the deposition of the main samples. The effective thickness of the \AlOx films varies only by a few per cent, AFM measurements showed variations of less then 2\,nm for a 20\,nm film thickness.
 
It should be pointed out that using the described technique, in spite of the very sharp transition of \AlOx from a low ohmic to an insulating state, see Fig.\ref{R_sq_vs_P}, we were able to reach the targeted sheet resistance with an uncertainty of better than $\pm$0.5\,k$\Omega$. This accuracy is sufficient for many applications employing highly resistive \AlOx.

\subsection{Heat treatment of \AlOx films}
\label{ThermalTreatment}
\begin{figure}[h!]
 \centering
 \includegraphics[scale=0.5]{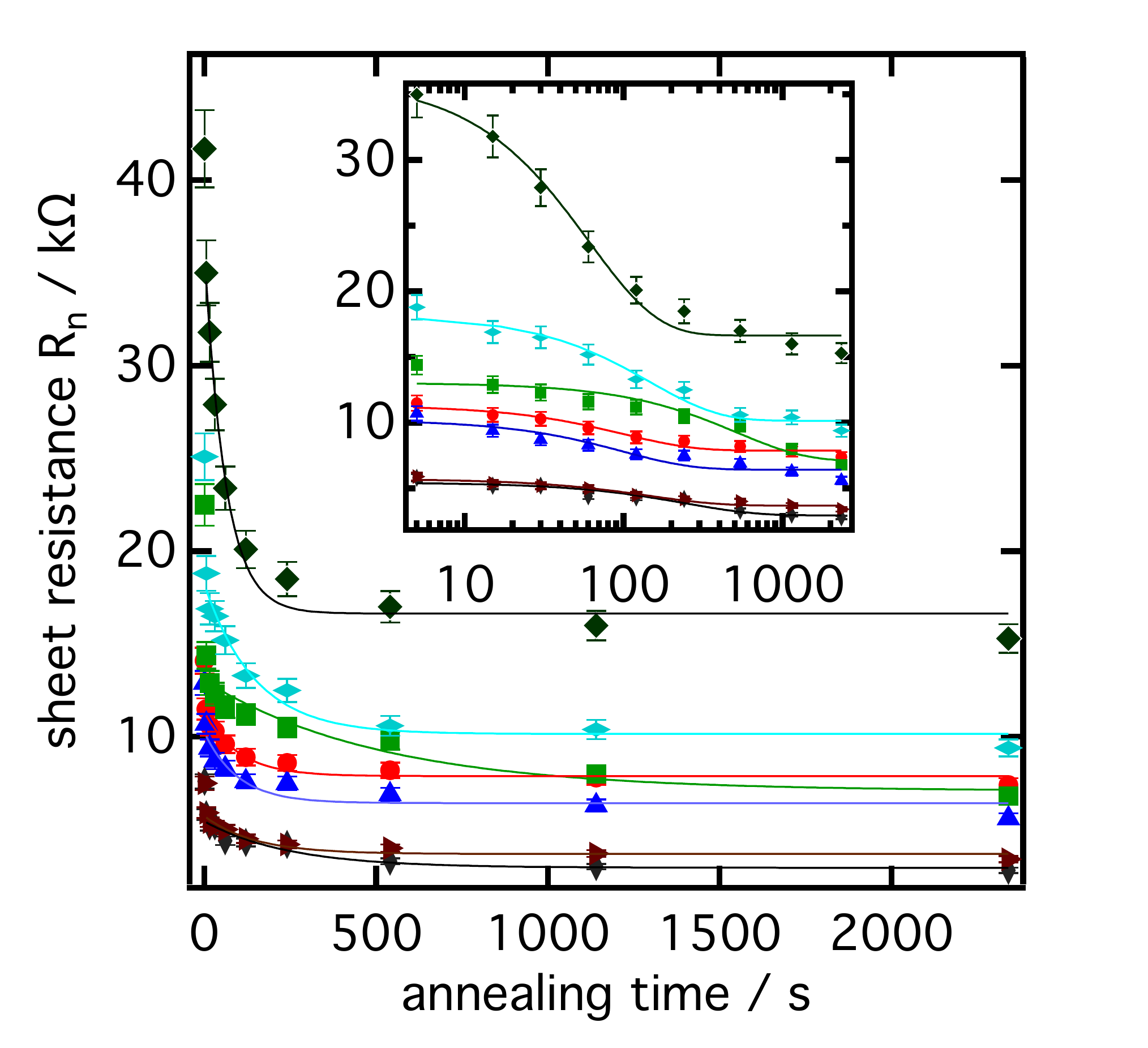}
 \caption{\label{R_ann_vs_time} Time dependence of the \AlOx sheet resistance at a fixed annealing temperature of 250\,$^\circ$C, overlaid are simple exponential decay curves as a guide to the eye. After $\approx$\,400-500\,s the sheet resistance enters a plateau value and does only change very slowly with increased annealing time. The inset shows the same data on a logarithmic time scale.}
\end{figure}

Patterning of \AlOx films by optical or e-beam lithography requires several steps during which certain heat treatments of the films are unavoidable. Heat treatment by itself  provides a convenient method of adjusting sample parameters after thin film processing. Therefore, we studied the dependence of $R_\mathrm{n}$ on thermal annealing for various \AlOx\ films grown on \Si\ and \Si\O$_2$ substrates. Similarly to the lithography steps, we annealed the films on a preheated hotplate for several minutes in air and measured the film resistance after the samples have been cooled down to room temperature. The same sample is annealed for a time or temperature step, measured and annealed again. The set of samples measured for the time dependent annealing experiment is different to that of the temperature dependent measurement.
Figure\,\ref{R_ann_vs_time} shows the annealing time dependence of $R_\mathrm{n}$ for 20\,nm thick \AlOx films annealed at a temperature of 250\,\Deg.  After a very steep decrease on the time scale of seconds, the sheet resistance $R_\mathrm{n}$ enters a plateau value at $400 - 500$\,s and changes only marginally for longer annealing times. A similar behavior of the mechanical properties has been observed for bulk aluminium and aluminium-alloy samples, though on a timescale of hours \cite{Ha84}. 

\begin{figure}[h!]
 \centering
 \includegraphics[scale=0.5]{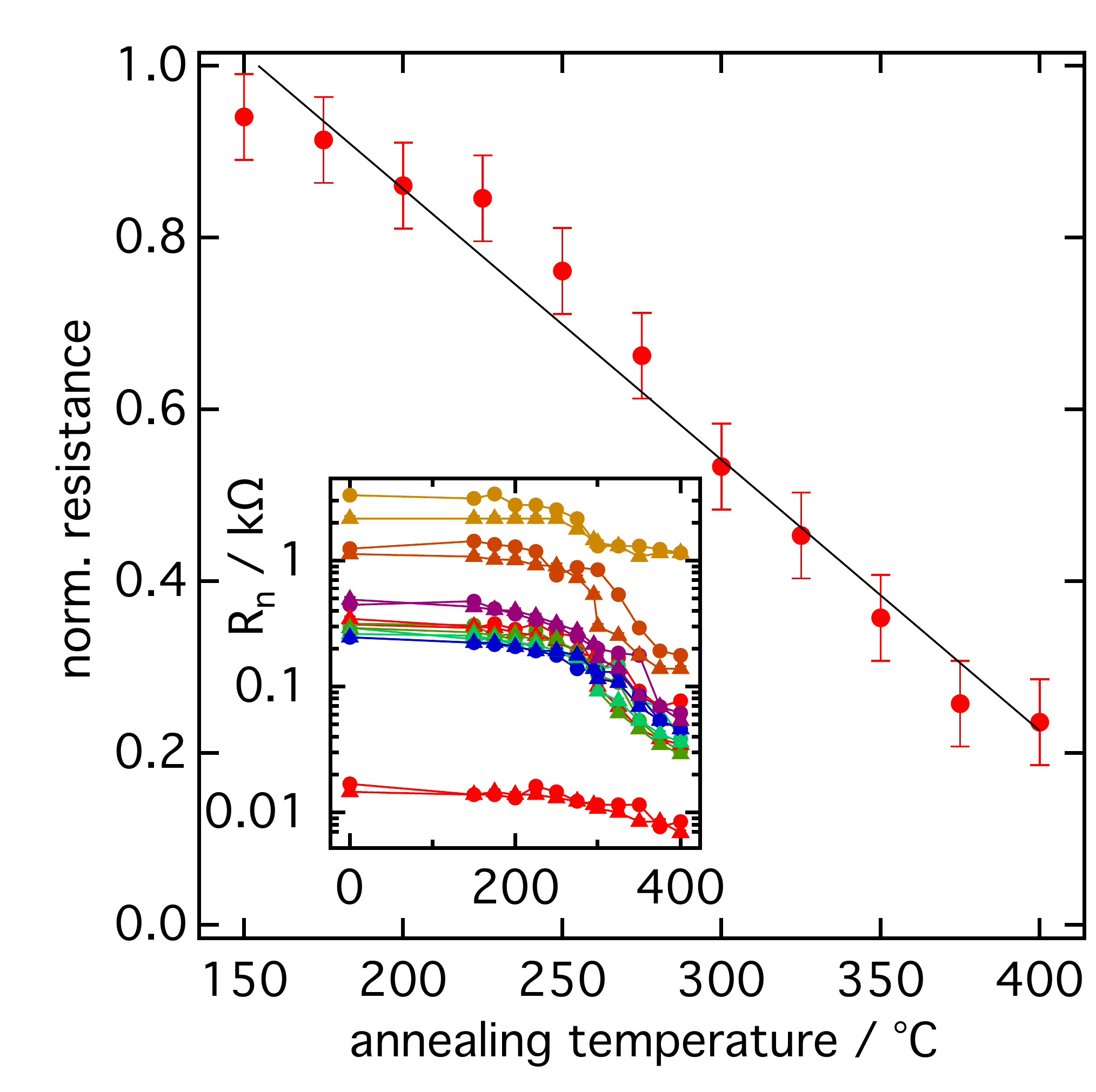}
 \caption{\label{R_ann_vs_temp} Normalized sheet resistance $R(T)/R(300\,K)$ of \AlOx films versus the annealing temperature after an annealing time of 600\,s. Shown is the average of individual films  (inset) grown on \Si\ and \Si\O$_2$ and having a wide range of initial sheet resistances.}
\end{figure}

Figure\,\ref{R_ann_vs_temp} shows the dependence of the normalized sheet resistance $R(T)/R(300\,K)$  of annealed \AlOx films grown on Si and SiO$_2$ and over a wide range of initial sheet resistances. Shown is the average value of all measured samples at a given temperature and annealing time of 600\,s. The inset of Fig.\,\ref{R_ann_vs_temp} contains the raw data as reference. When annealed below 200\,\Deg, the sheet resistance of the \AlOx thin films changes only by about 10\,\%, making this therefore fully compatible with conventional subsequent thin film patterning steps, i.e. the baking of lithographic resist. At 400\,$^\circ C$, the highest temperature that we have applied, we find a resistance reduction to about 22\% of the initial value. Between 200\,$^\circ C$ and 400\,$^\circ C$, the sheet resistance dependence with annealing temperature is found to be $R(T)/R(300\,K) \approx 1.5 - 3.2\times10^{-3}\,T /^\circ C$. This dependence is common to all samples, grown on \Si\ or \Si\O$_2$ substrates, having $R_\mathrm{n}$ below $R_\mathrm{n} \sim 20$\,k$\Omega$. Above this $R_\mathrm{n}$ value,  a fraction of the samples showed an increase of the resistance by heat treatment.

The un-annealed samples do not show any signs of the intrinsic granularity of the \AlOx films, measured by AFM scans (not shown). On the contrary, the rms surface roughness parameter \emph{R$_\mathrm{q}$} for most of the un-annealed samples is well below 0.5\,nm and is independent of the used substrates, \Si\O$_2$\  or \Si. In comparison, thermal or sputter grown pure aluminium films usually show \emph{R$_\mathrm{q}$} values of around 1\,nm or larger, due to the much larger grain size.

Concluding this section, we note a very low $R_\mathrm{n}$ drift $<5$\,\%, in general towards higher $R_\mathrm{n}$, at ambient storage temperatures measured over a period longer than one year.

\section{Microwave measurements}
\label{cpw_measurement}
\subsection{Experimental setup}
\label{setup}
\begin{figure}[h!]
 \centering
 \includegraphics[scale=0.75]{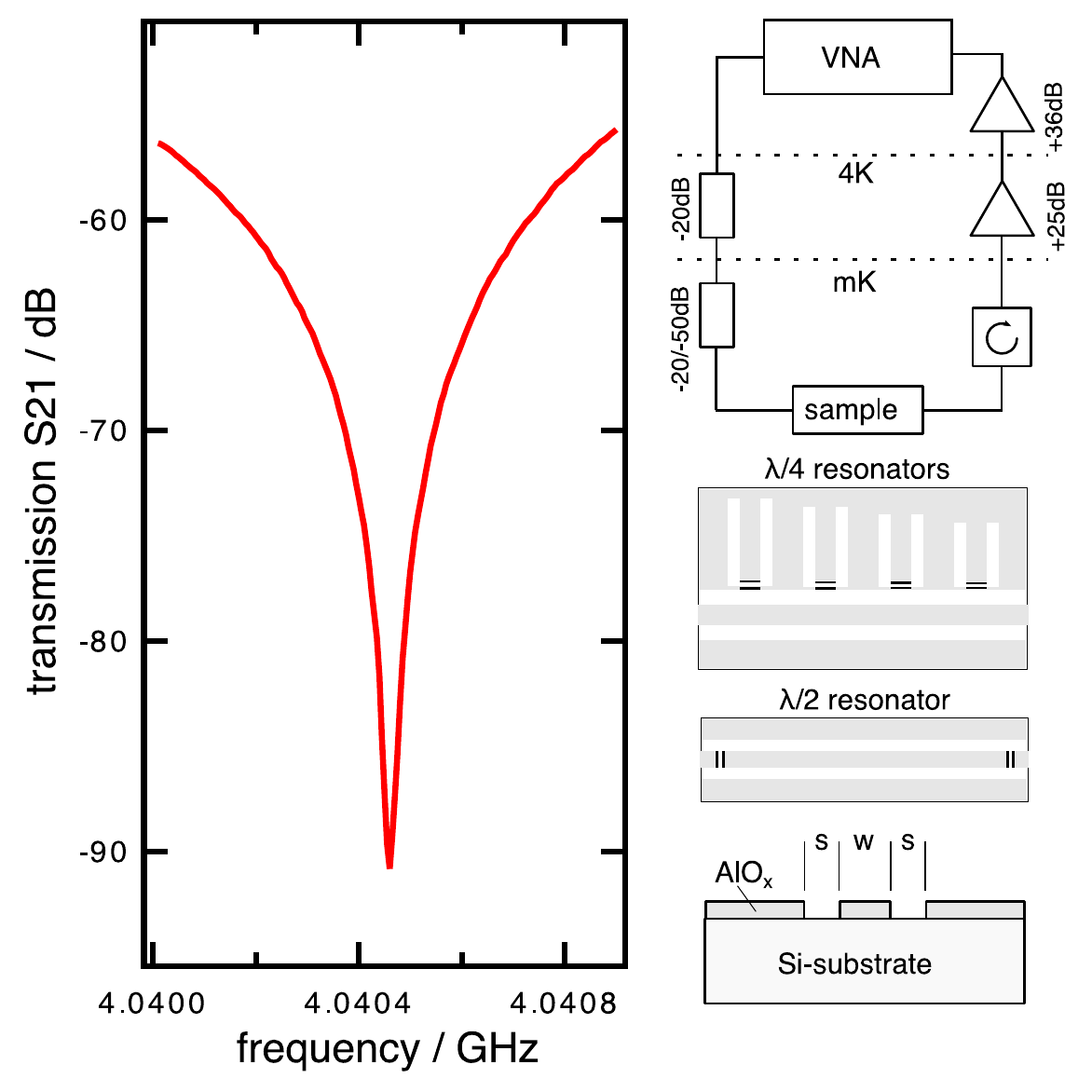}
 \caption{\label{setup_l4res} Microwave transmission measurement for a $\lambda$/4 resonator at -90\,dB power level (left) and a sketch of the experimental setup including the CPW resonators (right).}  
\end{figure}

We have measured the transmission of $\lambda/2$ and feedline coupled $\lambda/4$ coplanar waveguide (CPW) resonators made of grown \AlOx films using a $^3$He cryostat at $T\approx 300$\,mK and a cryogen free dilution refrigerator at $T\approx20$\,mK.  The measurement scheme was very similar for both setups and is sketched in Fig.\,\ref{setup_l4res} (right). The microwave probe signal from a vector network analyzer (VNA) was attenuated at different temperature stages by 40\,dB ($^3$He) and by 70\,dB (dilution refrigerator). In both setups, after passing the sample, the signal is amplified by 25\,dB by a cryogenic HEMT amplifier. In the dilution refrigerator, an additional isolator at sample temperature reduces the back action of the amplifier onto the sample.  At room temperature, a secondary amplifier boosts the signal by 36\,dB before it is fed into the VNA. 
The sample is enclosed in a copper EM-tight microwave sample box, connecting it to semi-rigid microwave cables via a 50\,$\Omega$ matched low-loss PCB.

\subsection{Resonator samples}
The samples were prepared by the deposition of a  20\,nm thick \AlOx film having a sheet resistance $R_\mathrm{n}$ = 0.5\,k$\Omega$ ($\lambda/2$ resonator) and $R_\mathrm{n}$ = 1.5\,k$\Omega$ ($\lambda/4$ resonator) on an intrinsic \Si\ substrate. The resist curing at 110\,$^\circ C$ for 50\,s and subsequent RIE etch did not change $R_\mathrm{n}$ significantly. We measured a residual resistivity ratio (RRR)  between room temperature and 4.2\,K of 1.3 for both films.

\Table{
\label{tabl_Qi} Internal quality factor $Q_i$ of a \AlOx $\lambda/4$ resonator measured at different power levels, at a resonance frequency $\omega_0 = 4.04$\,GHz. }
\br
Power / dBm & -80 &  -90 & -100 & -115 & -130 \\
$Q_i$ / $10^3$ & 343  & 668  & 674  & 329  & 126 \\ 
\br
\end{tabular}
\end{indented}
\end{table}

\subsection{Measurements at microwave frequencies}

To measure the internal quality factor $Q_i = (Q^{-1} - Q_c^{-1})^{-1}$, we have designed and fabricated a set of $\lambda/4$ resonators with different couplings to a common feedline. $Q$ is the total measured quality factor and $Q_c$ the coupling quality factor due to the coupling capacitors. Figure\,\ref{setup_l4res} shows a typical resonance curve measured at 4.04\,GHz frequency at -90\,dB power level. For the designed  $Q_c \approx 100\,000$, we measure internal quality factors ranging from 674 000 at intermediate to 126\,000 at low (single photon) power levels, see Table \ref{tabl_Qi}.
 
The measured $Q_i$ values show that the disordered \AlOx films seem to have only very low internal dissipation. Without going into more detail, it is feasible that the losses are dominated by the usual dielectric losses seen in many recent CPW resonator measurements, e.g. as in Ref.~\cite{Ga08,Sc15}. 

\begin{figure}[h!]
 \centering
 \includegraphics[scale=0.5]{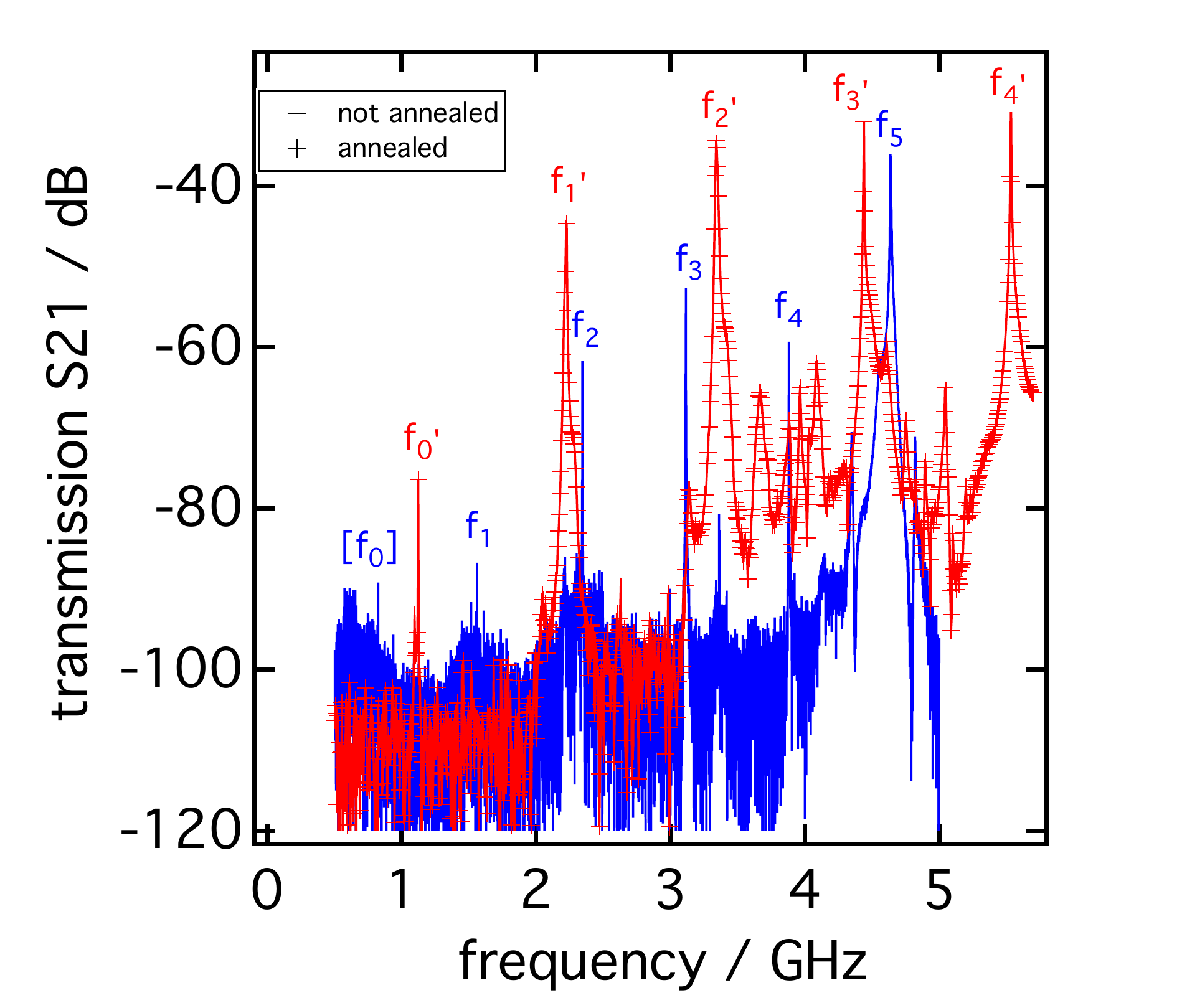}
 \caption{\label{S21_comparison} Comparison of the $\lambda/2$ CPW \AlOx resonator spectra measured before (blue dots) and after (red crosses) heat treatment at 250$\,^\circ$C for 10\,min. Several harmonic resonances ($f_n=n f_0$ for the $n$th harmonic frequency)  are visible in the measured frequency range. The fundamental mode $f_0$ of the un-annealed resonator (measured at $T=20$\,mK) is suppressed by the low frequency cut-off of the setup, but can be determined as $f_0 = f_{n}-f_{n-1}$. Due to the annealing, the sheet resistance and therefore the kinetic inductance both drop and $f_0$ is increased from 0.78\,GHz to 1.12\,GHz by $340$\,MHz, see Table \ref{tabl1} for details.}  
\end{figure}

For determination of the kinetic inductance, we focus on the measurement of the $\lambda/2$ resonator. This resonator was measured twice, first without heat treatment, in a dilution refrigerator at 20\,mK, and a second time after an annealing step at 250$\,^\circ$C for 10\,min in the $^3$He setup at about 300\,mK 
\setcounter{footnote}{0}
\footnote{The relative frequency shift due to a change of the kinetic inductance between 20\,mK and 300\,mK is less than $10^{-3}$, and can therefore be neglected.}. Figure\,\ref{S21_comparison} shows an overlaid comparison of the two measurements, solid blue is the initial spectrum and the red crosses are the data after the heat treatment.  Several harmonic frequencies of the resonator are clearly visible, the signal of the fundamental mode of both curves is suppressed by the lower cut-off frequency  ($\sim$2GHz) of the HEMT amplifiers. We find the base frequency by averaging the frequency spacing of the $n$th harmonic by $f_0 = f_{n}-f_{n-1}$, see Table \ref{tabl1} for the obtained values. The internal quality factor of the over-coupled ($Q_l \approx 2000$) resonator at the third harmonic frequency evaluates to $Q_i \approx 37\,000$ at 20\,mK.
After annealing, the room temperature sheet resistance was lowered from 497\,$\Omega$ to 266\,$\Omega$.

\Table{
\label{tabl1} \AlOx $\lambda/2$  resonator properties}
\br
Sample& R$_\mathrm{n}$$^{\rm a}$&L$_\mathrm{kin}$$^{\rm b}$&T$_\mathrm{c}$&$f_0$&$f_1$&$f_2$&$f_3$&$f_4$&$f_5$\\
            &$\Omega$/sq & pH/sq & K & GHz & GHz &  GHz &  GHz &  GHz &  GHz \\ 
\mr
un-annealed       & 382 & 329 & 1.6 & [0.78] & 1.56 & 2.34 & 3.11 & 3.87 & 4.65 \\
annealed             & 205 & 176 & 1.6 & 1.12   & 2.23 & 3.35 & 4.44 & 5.53 &\\
\br
\end{tabular}
\item[] $^{\rm a}$ at 4.2\,K, RRR = 1.3.
\item[] $^{\rm b}$ see relation (\ref{Lkin_R}).
\end{indented}
\end{table}

\subsection{Discussion}\label{lambda_2}

From a general point of view, the thin film resonator is a quasi two dimensional structure since its thickness $d = 20$\,nm is of the order of the effective coherence length $\xi = \sqrt {\xi_0 \, l}$. Following Ref.~\cite{De73},  $\xi$ is of the order of the grain size $\approx 4$\,nm. An estimate of the penetration depth $\lambda_\mathrm{L}$ at $T \ll T_\mathrm{c}$ is given by $1.05\times10^{-3} \sqrt{R_\mathrm{n}\,d/ T_\mathrm{c}}$ m \cite{Yo95}, for which we find $\lambda_\mathrm{L} \approx$ 2.2\,$\mu$m (un-annealed) and $\lambda_\mathrm{L} \approx 1.6\,\mu$m (annealed). Therefore $\lambda_\mathrm{L}$ is much larger than the film thickness and also the perpendicular penetration depth $\lambda_\mathrm{perp} = \lambda_\mathrm{L}^2/d$ is much larger than the center conductor width.

The \AlOx resonators were fabricated on intrinsic silicon ($\epsilon_r = 11.7$), therefore $\epsilon_\mathrm{eff}$ is 6.3 \cite{Si01}. With $w=10\,\mu$m and $s = 6\,\mu$m, see Fig.\,\ref{setup_l4res},  $L_m$ and $C$ are evaluated to be $438$\,nH\,m$^{-1}$ and $160$\,pF\,m$^{-1}$, respectively. 
Using relation (\ref{freq_LC}), we find a base resonance frequency of $6.7$\,GHz for the resonator with the length $8.96$\,mm, neglecting the kinetic inductance for a moment. This is reasonably close to a detailed high frequency EM finite element simulation data 
\renewcommand{\thefootnote}{$\star\star$}
\footnote{Sonnet Software, Inc., em Version 13.56, 100 Elwood Davis Road, North Syracuse, NY 13212, USA, 2011}, which in addition  includes also the geometrical details of the coupling capacitors. The EM solver calculates $f_0 =  6.9$\,GHz as the base resonance frequency, thus the influence of the coupling capacitor contributes to about  3\,\%  in $f_0$ and is neglected in the further discussion. 

\begin{figure}[h!]
 \centering
 \includegraphics[scale=0.5]{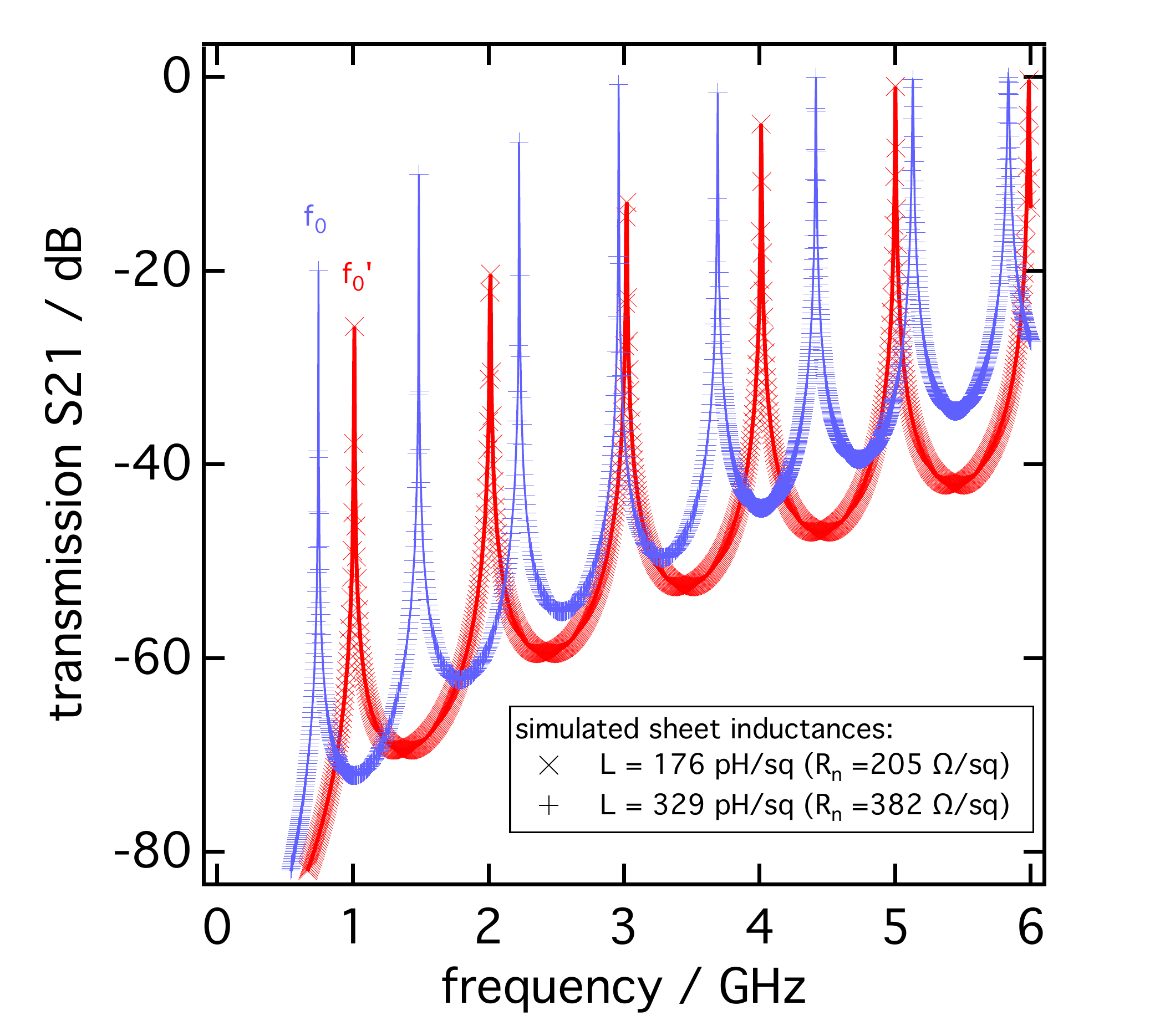}
 \caption{\label{EM-sim} EM-simulation results of the $\lambda/2$-resonator with a sheet inductance 176\,pH/sq (red, $\times$) and 329\,ph/sq (blue, +). The sheet inductance was derived from the measured $R_\mathrm{n}$ value and Eq.~(\ref{Lkin_R}). The simulated fundamental frequencies deviate from the measured frequencies by less than 10\,\%, see also Fig.\,\ref{S21_comparison}.}
\end{figure}

If a kinetic sheet inductance derived from $R_\mathrm{n}$ and Eq.~(\ref{Lkin_R}) is included in the EM-simulation, we obtain base frequencies of  745\,MHz (un-annealed) and 1.01\,GHz (annealed), see Fig.\,\ref{EM-sim}  for a spectrum from 0 to 6\,GHz of the simulated S21 transmission. 
In the EM-simulation, the center conductor and ground planes are treated as ideal conductors with the chosen material inductance to be the equivalent to the kinetic inductance corresponding to $R_\mathrm{n}$, with no further "superconducting" effects included. The calculation grid was set to $2\,\mu$m.
Compared with the measured base frequencies the agreement is, likely due to the large $\lambda_\mathrm{L}$, rather good,  the EM-solver result overestimates the influence of the additional inductance by around 6\,\% (un-annealed) and 10\,\% (annealed).
 
In the analytic calculation, the high frequency pre-factor to the kinetic inductance $\alpha = g_\mathrm{kin} \, w$ contains the details of the current distribution over the resonator cross section, which, in principle, also depends on the material's inductance.
Instead of investigating the details of the current distribution in the centre conductor cross section or temperature dependent measurements of $\lambda_\mathrm{L}$, as treated e.g. in Ref.\,\cite{Po95}, we obtain $\alpha$ as a free parameter from the measured base frequency $f_0$ and $R_n$.  With Eq.~(\ref{Lkin_R}) and  $\alpha =  w L_\mathrm{kin}^{-1}[((2 f_0 l )^2 C)^{-1} - L_m]$, we find $\alpha = 0.96$ (un-annealed) and $\alpha = 0.86$ (annealed).
Using the same method for comparison, we find from the EM-simulation data a constant value of $\alpha = 1.06$ for both $R_\mathrm{n}$ values. 

It is interesting to note, that due to the large total inductance  $L_\mathrm{tot}=1/(2f_0 l)^2 C =286.4\,$nH (un-annealed) and  $L_\mathrm{tot}=138.7\,$nH (annealed), we find a characteristic resonator impedance $Z_0 = \sqrt{L/C} = \sqrt{1/(2 f_0 l)^2 C^2}  = 447$\,$\Omega$ (un-annealed) and $Z_0 = 311\,\Omega$ (annealed). These high resonator impedance values may be an interesting property for microwave kinetic inductance detectors (MKID), because it can be easily matched e.g. to the vacuum impedance $Z_{0,\mathrm{vac}} = \sqrt{\mu_0/\epsilon_0} = 377\,\Omega$ \cite{Mo11}. Compared with the total magnetic inductance $L_\mathrm{m,tot} = 3.9$nH, we find a kinetic inductance fraction \cite{Ga06} $\alpha_\mathrm{kin} = L_\mathrm{kin,tot}/(L_\mathrm{kin,tot} + L_\mathrm{m,tot} ) = 0.99$ and $0.97$.

While it is possible to estimate the base frequency of a high kinetic inductance resonator by an EM-simulation, as we have shown above,  we want to point out a less numerically intensive way of estimating a frequency shift by using Eq.\,(\ref{Lkin_R}).\\
The comparison of the values for the base frequency before ($f_0$) and after annealing ($f_0'$), can be written using relation (\ref{freq_LC})  to
$f_0/f_0' = ((L_m+g_\mathrm{kin}'L_\mathrm{kin}')/(L_m+g_\mathrm{kin} L_\mathrm{kin}))^{1/2}. $
For resonators with a high kinetic inductance, $L_m \ll g_\mathrm{kin} L_\mathrm{kin}$, ($\alpha_\mathrm{kin} \approx 1$), the relation approximates with Eq. (\ref{Lkin_R}) to
\begin{equation}\label{f0f0_RnRn}
\frac{f_0}{f_0'} = \sqrt{\frac{L_\mathrm{kin}'}{L_\mathrm{kin}}} = \sqrt{ \frac{T_c g_\mathrm{kin}' R_n'}{T_c' g_\mathrm{kin} R_n}}.
\end{equation}
The $T_c$,  $f$ and $R_n$ values are measured independently. In our case, $T_c = 1.6$\,K was measured to be the same for both resonator measurements. For the $g_\mathrm{kin}$ change $g_\mathrm{kin}' = k\,g_\mathrm{kin}$ due to thermal annealing, we find with $k = (f_0/f_0')^2R_n/R_n'$ a value of $k= 0.9$. We can therefore simplify Eq.~(\ref{f0f0_RnRn}) within an error margin of about 5\,\% to $f_0/f_0' \approx \sqrt{R_n'/R_n}$. 
This result is of practical importance, because it allows to estimate a resonator base frequency by a simple sheet resistance measurement after thermal annealing. 

\section{Conclusion}
In this paper, we show that by adding oxygen impurities to aluminium thin films, wires with a widely adjustable kinetic inductance at moderate film thicknesses can be obtained. Fabricated with a controlled DC magnetron sputter growth process, we found robust conditions for room temperature sheet resistances in a range from $0.1$ to several \,k$\Omega$, corresponding to a high kinetic wire inductance in the superconducting state. As an option for lowering the sheet resistance, the \AlOx films can be annealed at moderate temperatures, $R_\mathrm{n}$ drops to  about $1/5$ of the initial value with a heat treatment at 400\,$^\circ$C. 

Measurements of the microwave response of superconducting resonators confirm the high kinetic inductance. Moreover, the results demonstrate the potential of \AlOx to serve as a low-loss and high $Q$ resonator material e.g. for microwave kinetic inductance detectors or compact resonators used, for instance, with superconducting qubit circuits.

\section{Aknowlegements}
We thank M. Dries and D. Gerthsen for the help with the TEM imaging, L. Radtke for general support in the clean room facilities. This work was supported by the DFG  Center for Functional Nano-structures (CFN) Karlsruhe and the DFG Research Unit 960 Quantum Phase Transitions. S.T. S. acknowledges support from the Heinrich B\"oll Stiftung.

\section*{References}

\bibliography{citations/references}{}
\bibliographystyle{unsrt}

\pagebreak

%\section{Supplementary material}
%\widetext
\begin{center}
\textbf{\large Supplemental Materials: Aluminium-oxide wires for superconducting high kinetic inductance circuits}
\end{center}

%%%%%%%%%% Prefix a "S" to all equations, figures, tables and reset the counter %%%%%%%%%%
\setcounter{equation}{0}
\setcounter{figure}{0}
\setcounter{table}{0}
\setcounter{page}{1}
\makeatletter
\renewcommand{\theequation}{S\arabic{equation}}
\renewcommand{\thefigure}{S\arabic{figure}}
\renewcommand{\thetable}{S\arabic{table}}
%\renewcommand{\bibnumfmt}[1]{[S#1]}
%\renewcommand{\citenumfont}[1]{S#1}
%%%%%%%%%% Prefix a "S" to all equations, figures, tables and reset the counter %%%%%%%%%%

\section{Sputter deposited granular \AlOx: TEM}
\begin{figure}[h!]
 \begin{center}
 \includegraphics[scale=0.5]{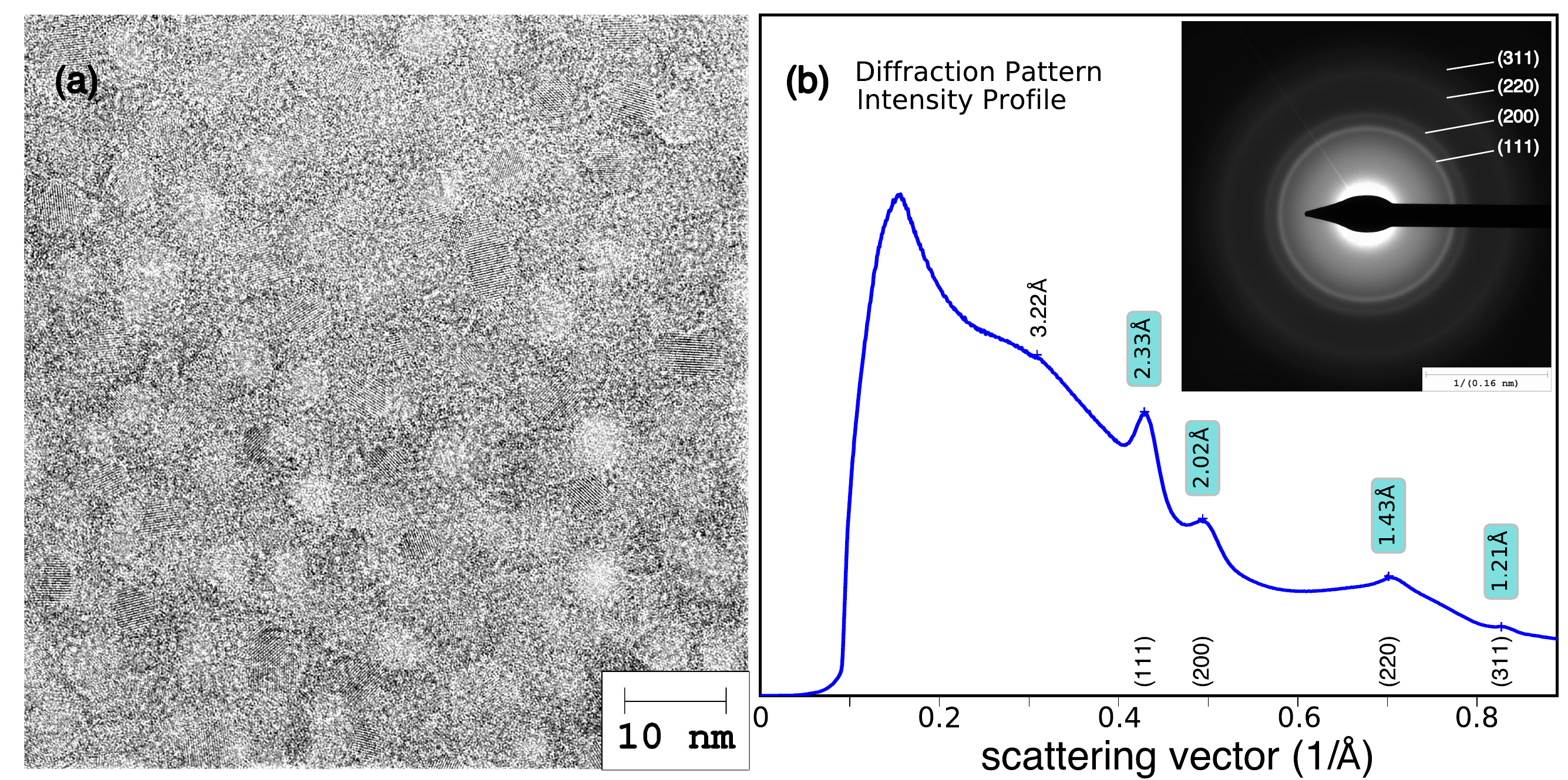}
 \caption{\label{TEM}(a) TEM image of a 20\,nm thick \AlOx film with a sheet resistance of about 2\,k$\Omega$, without heat treatment. The individual grains have a mean size of about 4\,nm and a random orientation.  (b) Intensity profile and TEM diffraction pattern of the same film. The peaks with the colorised labels are in agreement with the corresponding lattice planes of the face-centered cubic lattice of bulk aluminium.}
\end{center}
\end{figure}

The granularity of the sputter deposited \AlOx films was determined by a transmission electron microscopy (TEM) measurement, see Fig.\ref{TEM} (a). 
20\,nm thick films have been sputter deposited on top of a thin soap film on a mica substrate. Before lifting and transferring  to a TEM mesh, we determined a sheet resistance of about 2\,k$\Omega$ for the shown film. 

Thermally evaporated granular aluminium films have been studied extensively and similar images have been reported, e.g. in Ref.~\cite{De73}. The TEM image acquired from the sputter deposited samples in this work show a polycrystalline film with randomly orientated nano-crystallites and an average grain size of about 4\,nm. The orientation and internal structure of the grains can be deduced from a TEM diffraction pattern (Fig.\,\ref{TEM}(b)), taken from the same film. The concentric rings identify the face-centered cubic  structure of bulk aluminium with the bulk aluminium lattice parameter of $a=0.40415$\,nm \cite{St49}.

Except for a faint peak at $3.22$\AA, we do not observe rings which cannot be attributed to pure aluminium, thus there is no clear indication how the oxygen is distributed in the film. In comparison, $\alpha-\Al_2\O_3$  has a trigonal crystal structure with the lattice parameters $a = 0.4785$\,nm and $c = 1.2991$\,nm.  Table \ref{TEMtable} lists the the measured lattice spacings with simliar bulk lattice spacings of $\Al$ and $\Al_2\O_3$ \cite{St49,Do09}:

\begin{table}[htp]
\caption{Measured lattice spacings compared with lattice spacings taken from \cite{St49,Do09} for $\Al$ and $\alpha-\Al_2\O_3$. All distances are in \AA\ (0.1nm). The uncertainty on the measured values are in the order of $\pm0.03$ \AA.}
\begin{center}
\begin{tabular}{|c||c|c||c|c|}
$d$(meas) & $d(\Al)$ & $hkl$ & $d(\Al_2\O_3)$ & $hkl$ \\
\hline
3.22 & - & - & 3.479 &  (012)\\
2.33 &  2.3333  &  (111) & 2.552, 2.379 & (104), (1110) \\
2.02 &  2.0207  &  (200) &  2.085 & (113) \\
1.43 & 1.4289   &  (220) & 1.374, 1.404 &(030), (124)\\
1.21 & 1.2185   &  (311) & - & -\\
\end{tabular}
\end{center}
\label{TEMtable}
\end{table}

Between the grains in Fig.\,\ref{TEM}(a) are regions which show no regular pattern, but apart from a disordered structure, this image parts could also stem from overlapping grains with different orientation, since the film is substantially thicker than the grain diameter (assuming spherical grains).

It is feasible, that the oxygen/aluminium-oxide is located on the surface of the aluminium grains\cite{De73,Ab76}, however the thickness and structure of such a layer we could not directly deduce. In the context of a Josephson junction network model, this would be an important parameter since it mediates the coupling energy between the grains.

\end{document}